\documentclass{aastex}          
\usepackage{spr-astr-addons}    

\begin{document}
%
\title{The evolution of star clusters:
The resolved-star approach}

\shorttitle{The resolved-star approach to star cluster evolution}
\shortauthors{Pellerin et al.}

\author{Anne Pellerin} 
\affil{Johns Hopkins University, Physics \& Astronomy Department, 3400 N. Charles Street, Baltimore, MD 21218, USA}
\author{Martin J. Meyer}
\affil{University of Western Australia, 35 Stirling Highway, Crawley WA 6009, Perth, Australia}
\author{Jason Harris}
\affil{Steward Observatory 933 N. Cherry Ave. Tucson, AZ 85721, USA}
\author{Daniela Calzetti}
\affil{University of Massachusetts, Department of Astronomy, 710 North Pleasant Street, Amherst, MA 01003, USA}

\begin{abstract}
We present the first results of a new technique to detect, locate, and
characterize young dissolving star clusters. Using {\sl Hubble Space
Telescope}/Advanced Camera for Surveys archival images of the nearby
galaxy IC2574, we performed stellar PSF photometry and selected the
most massive stars as our first test sample. We used a group-finding
algorithm on the selected massive stars to find cluster candidates. We
then plot the color-magnitude diagrams for each group, and use stellar
evolutionary models to estimate their age. So far, we found 79 groups
with ages of up to about 100 Myr, displaying various sizes and
densities.
\end{abstract}


%
\section{Introduction}
\label{intro}

Star clusters are fundamental building blocks of galaxies, since they
are the birth places of most stars \citep{lada03}. Despite tremendous
efforts to study their properties, the literature is relatively scarce
when it comes to describing observationally their evolutionary
sequence, and particularly the latest stages of disruption and
dissolution. The reason for this is simple: because of their low
surface brightness, they are very difficult to distinguish from the
main stellar background of their host galaxy. However, to better
understand the evolution of stellar clusters from birth to death, we
need to find a way to observe them even in their latest stages.

Recent studies suggest that the dissolution and disruption of star
clusters can happen at very young ages
\citep[e.g.][]{fall05,pel07}. Their work suggests that `infant
mortality' of star clusters might be responsible for the loss of 90\%
of the star clusters within the first 10\,Myr of their life. Infant
mortality of star clusters occurs when a cluster loses a significant
fraction of its gaseous mass during the supernovae explosion phase
\citep{lada03,bas06}, leaving the cluster gravitationally
unbound. Following those studies, young stars can be used to study
dissolving clusters, because cluster dissolution seems to commonly
happen at young ages. Also, massive stars are relatively easy to
observe and distinguish from the main stellar background of the host
galaxy due to their brightness and color. Our idea here is to use the
individual stellar properties to find groups of stars, including the
systems that are very dispersed and difficult to observed by means of
traditional integrated photometric techniques.

In this contribution, we present the first results of a new approach
to detect, locate, and characterize young dissolving star clusters
using the properties of resolved massive stars. In the next section we
describe the data and the details of the photometry. In Section\,3 we
explain the technique used to find groups of stars. The first results
are presented in Section\,4, including concrete examples of clusters
found in the nearby galaxy IC2574. Note that the term `star cluster'
will be use here to describe a wide range of stellar groups presumably
born together at approximatively the same time, and will include
gravitationally unbound systems as well as those very dispersed. It
may include associations, OB groups and open clusters.

\section{{\sl Hubble Space Telescope}/Advanced Camera for Surveys data and PSF photometry}
\label{phot}

To test our new approach to find dissolving star clusters, we need a
nearby galaxy for which the stellar content is spatially
resolved. Therefore, we used Cycle~12 archival data of the local
galaxy IC2574 observed with the {\sl Hubble Space Telescope (HST)}
using the Advanced Camera for Surveys (ACS) in the F435W, F555W, and
F814W filters (see Fig. 1). IC2574 is a spiral galaxy located at
4.02\,Mpc \citep{kara03}, which results in an ACS resolution limit of
2 pc, so that we are therefore capable of resolving individual
stars. The ACS field covers the southern half of the visible galaxy,
but excludes the nuclear region. Fig. 1 shows important star-forming
regions, which are mainly seen in two complexes and associated with
nebular emission. The data were reduced with the STScI calibration
pipeline, which includes treatment for geometric correction, cosmic
rays and drizzling.

\begin{figure}[t]
\includegraphics[width=\columnwidth]{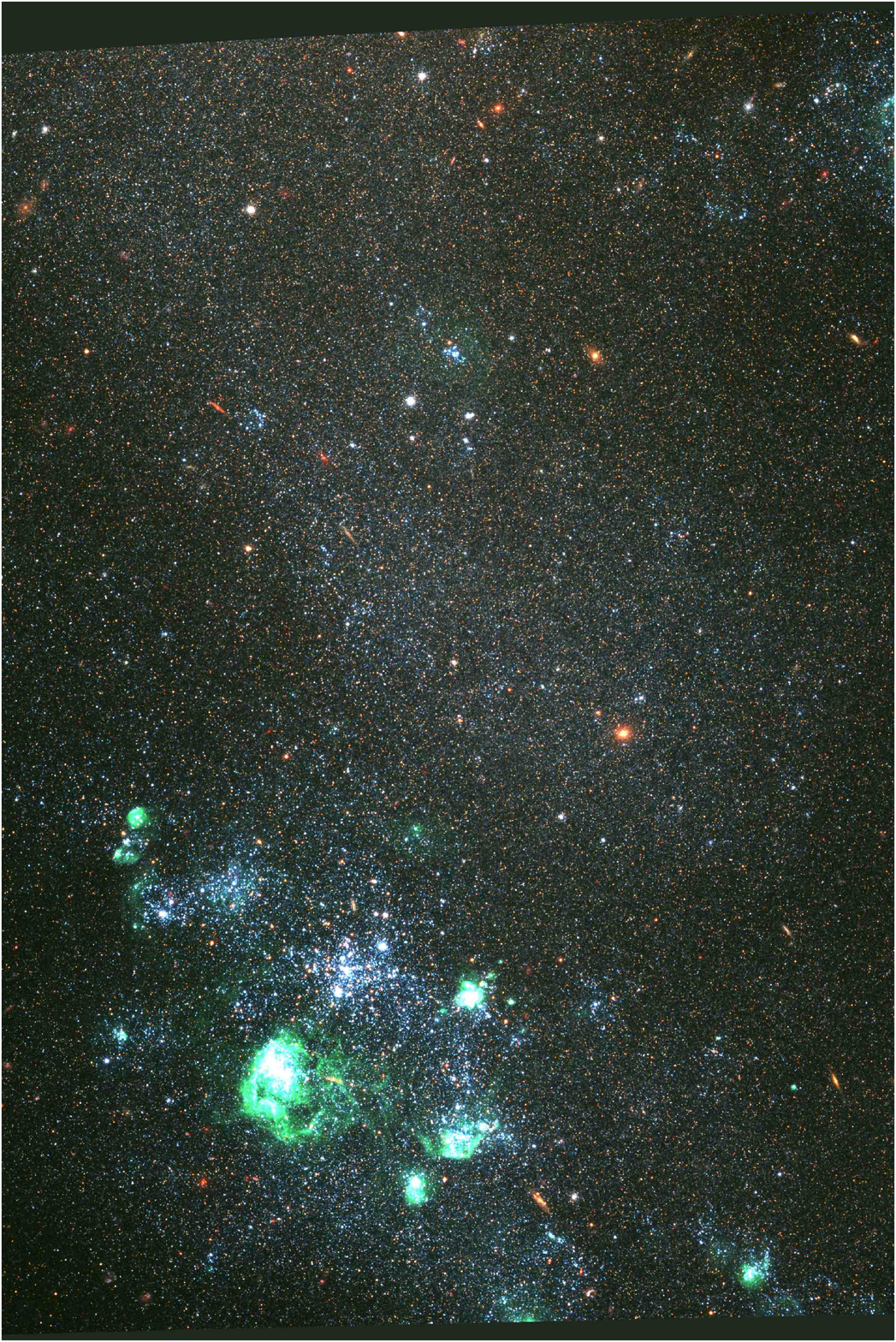}
\caption{Three-color image of IC2574 observed with {\sl HST}/ACS in the
F435W (blue), F555W (green), and F814W (red) filters. North is up}
\label{ic2574}
\end{figure}

We needed to perform point-spread function (PSF) photometry on each
image because of the crowding. We used the {\sc iraf/daophot} package
and about 100 unsaturated bright stars to create a PSF model with an
average FWHM of 2\,pc. We used the ACS photometric zero points of
\citet{sir05} in the ST system. We applied a homogeneous foreground
Galactic extinction, $E(B-V)=0.036$ mag (Schlegel et al. 1998) as well
as aperture corrections. The 50\% completeness levels are estimated at
27.7, 26.8, and 28.7\,mag for the F435W, F555W, and F814W filters,
respectively. A detailed color-magnitude diagram (CMD) of IC2574 is
presented in Fig. 2.

\begin{figure}[t]
\includegraphics[width=\columnwidth]{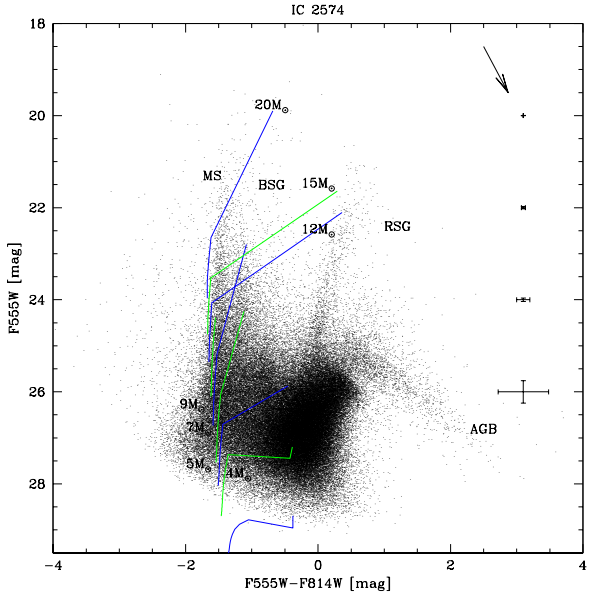}
\caption{Color-magnitude diagram of IC2574 obtained from the {\sl
HST}/ACS images in the F555W and F814W filters. Stellar evolutionary
tracks from 4 to 20M$_{\odot}$ are from \citet{gir06}. Typical
photometric uncertainties are shown on the right, together with the
1\,mag extinction vector}
\label{ic2574}
\end{figure}

\section{The resolved-star approach}
\label{method}

The main idea of the resolved-star approach is to use individual stars
to recreate the stellar clusters, even those that do not contrast well
with respect to the stellar background. The first step of the method
is, therefore, to obtain the photometric information on the individual
stars contained within a relatively large field ($\geq$1 kpc) of a
galaxy, i.e., IC2574 in our test case. The large field is required to
make sure that we detect the most extended clusters. For the reasons
mentioned in Section\,1, we selected the massive stars on the main
sequence (i.e., $\mbox{F555W-F814W} < 1.0$ mag) and brighter than
25.0\,mag in the F555W filter (see Fig.\,2). We did not use fainter
stars to avoid excessive photometric uncertainties, and also to make
sure that the stars contrast well with respect to the stellar
background of the host galaxy.

The second step, and the key to this approach, is the use of a
group-finding algorithm to find groups of stars within the selected
sample of massive stars. We chose the `HOP' algorithm \citep{hop}, a
clustering algorithm traditionally used in cosmology. HOP is a
sophisticated version of friends-of-friends algorithms; it is based on
the spatial density of stars instead of the distance between stars. In
other words, the algorithm 'hops' each particle (or star) to its
neighbor having the highest density, instead of the nearest
neighbor. It also allows us to specify useful parameters such as a
density value to be used to separate groups, a maximum density value
for a group to be viable and not merged, and the minimum density value
for a particle to be member of a group. These parameters basically
allow us to find groups within groups.

Because there are several free parameters involved in HOP, the
algorithm gives multiple solutions. Consequently, we considered that a
group found with HOP was a viable stellar group when the same group
was found using multiple sets of parameters.

The third step is to study the characteristics of the stellar
population of the stellar group. Since we selected only the most
massive stars to run the clustering algorithm, studying only the stars
in the HOP group would obviously lead to an incomplete result, because
of the lack of the less massive stars and of evolved stars such as the
red supergiants.

Therefore, for each group found with HOP, we go back to the ACS images
and create a region around the HOP group that includes all of the
detected stars, to estimate the size of the group. The size of the
region is defined by the stars contained within a radius, $r$, around
each massive star used to run HOP, where $r$ is a fraction (from 10\%
to 90\%) of the distance between the two most distant stars of the
given HOP group. This simple technique has two advantages. First, it
keeps the information on the shape of the HOP group. Secondly, it
allows us to find how extended the group is by comparing the CMD of
the group for various values of $r$. For each increasing step of $r$,
stars are added to the group. When the stars added to the group do no
longer contribute to the brightest regions of the color-magnitude
diagram of the group, the added stars are considered part of the
stellar background, and the size of the group is defined. Ultimately,
we compare the CMD diagram to Padova isochrones \citep{gir06} for each
group. Despite the fact that some groups can have a relatively small
number of stars, an estimate of the age and mass can be obtained for
each group.  A few concrete examples of the technique are given in the
next section.

\section{Cluster candidates}
\label{clus}

To test our new approach, we applied it to the stars observed in the
{\sl HST}/ACS field of IC2574. As expected, the HOP algorithm gave
different sets of outputs. An example of the output is shown in
Fig. 3. In this specific case, the parameter values used to find
groups produced the merging of the densest groups within the two big
complexes, but allowed us to separate the most dispersed groups from
each other.

After examining several HOP outputs, we found a large variety of
stellar groups outside of the two young complexes. We excluded the two
young-complex regions to test our approach on dissolving clusters,
which are much more difficult to identify than the young and compact
groups of the complexes. In the following, we present a few stellar
groups found using HOP, with various physical properties.

\begin{figure}[t]

\includegraphics[width=\columnwidth]{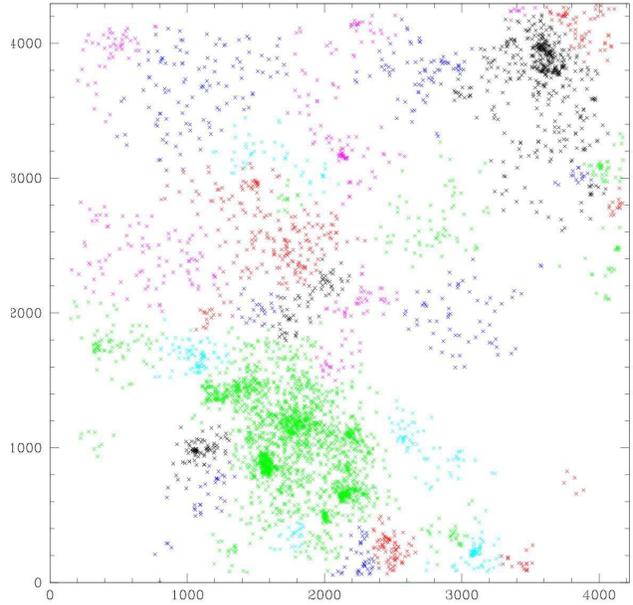}
\caption{Example of an HOP output for the massive stars in
IC2574. Various colors have been used to represent individual
stellar groups. The axes corresponds to the ACS pixels, and the scale
is about 1\,pc pixel$^{-1}$ for this galaxy}
\label{hopgroup}
\end{figure}

\subsection{Young and compact cluster with nebular emission} 

In Fig. 4, we show a small section of Fig.\,1 which zooms in on a
young and compact group found with HOP. From the nebular emission, it
is obvious that the group contains hot OB stars ionizing its
surrounding gas.

Fig. 5 shows the CMD of the group, which does not include any evolved
star younger than about 100\,Myr. Also, bright massive stars are seen
on the main sequence, consistent with a 30\,Myr isochrone, or
younger. Because of the nebular emission, we estimate the age of this
group to be less than 10\,Myr.

Note the presence of the dashed square in the bottom right-hand corner
of the CMD. This represents a region where evolved stars, such as red
giants, are found. We assume this region to be filled with background
stars, or at least with stars too old to be related to the young
stellar group. To be in the dashed area of the CMD, the stars must be
as old as $\sim$100\,Myr, or older. Thus, we can see that the CMD of
even young star clusters can be contaminated by a significant number
of background stars. In this case, it is likely that a large fraction
of the contaminatin stars are not spatially coincident with the
cluster, but simply observed in the same field of view.

\begin{figure}[t]
\includegraphics[width=\columnwidth]{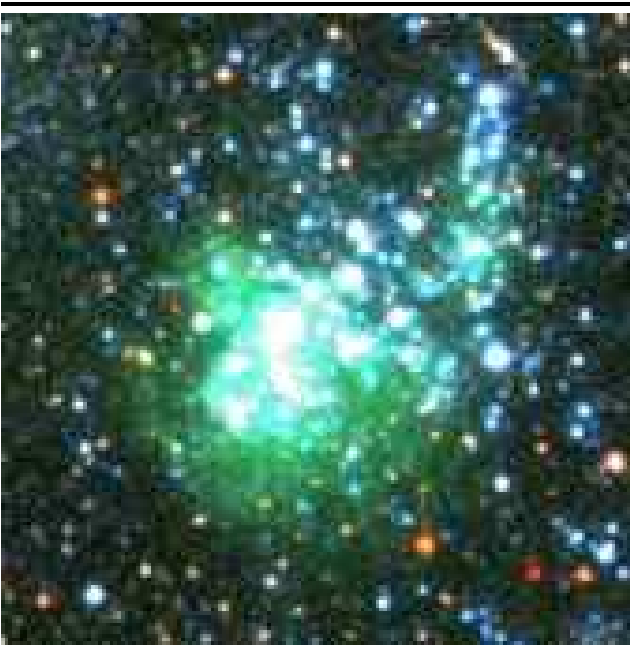}
\caption{Example of a young and compact group found with HOP. The
image shows a section of about 7\,pc from Fig. 1}
\label{imneb}
\end{figure}

\begin{figure}[t]
\includegraphics[width=\columnwidth]{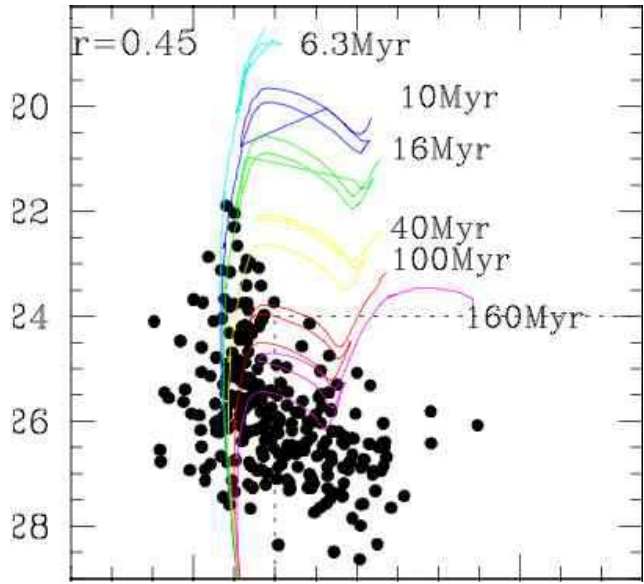}
\caption{CMD of the stellar group shown in Fig.\,4. The axes are as in
Fig.\,2}
\label{cmdneb}
\end{figure}

\subsection{Young cluster without nebular emission}

Fig. 6 shows a stellar group of massive stars with no nebular
emission. The group is relatively compact, but dispersed enough so
that the stars can be resolved easily using PSF photometry. Note that
the cluster is about twice the size of the previous cluster (Section
4.1). The CMD of the group, displayed in Fig. 7, shows several stars
that are well fitted by the 40\,Myr isochrone. However, one object is
clearly brighter, and would be fitted much better by younger
isochrones. One must not forget that we are dealing with smal-number
statistics, and that the bright object is possibly a multiple system,
or even a peculiar star. It might also be a runaway star from a nearby
system. All we can say for sure is that this group of stars is
probably about 40\,Myr or younger.

\begin{figure}[t]
\includegraphics[width=\columnwidth]{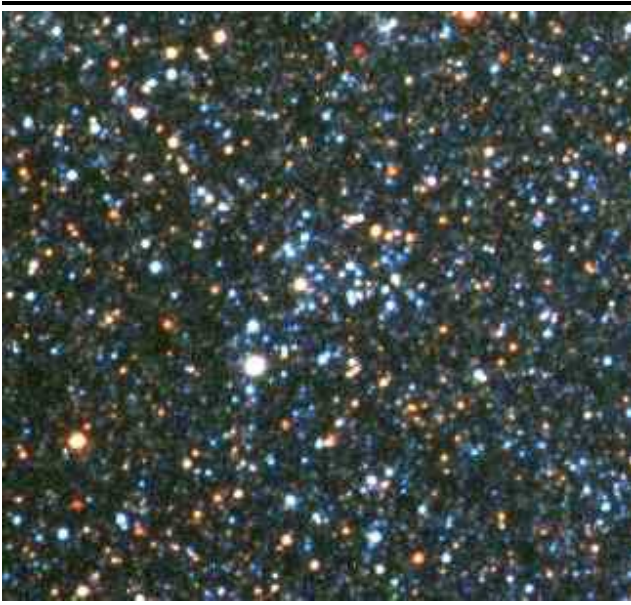}
\caption{Example of a young group without nebular emission. The size is about 15\,pc} 
\label{imnoneb}
\end{figure}

\begin{figure}[t]
\includegraphics[width=\columnwidth]{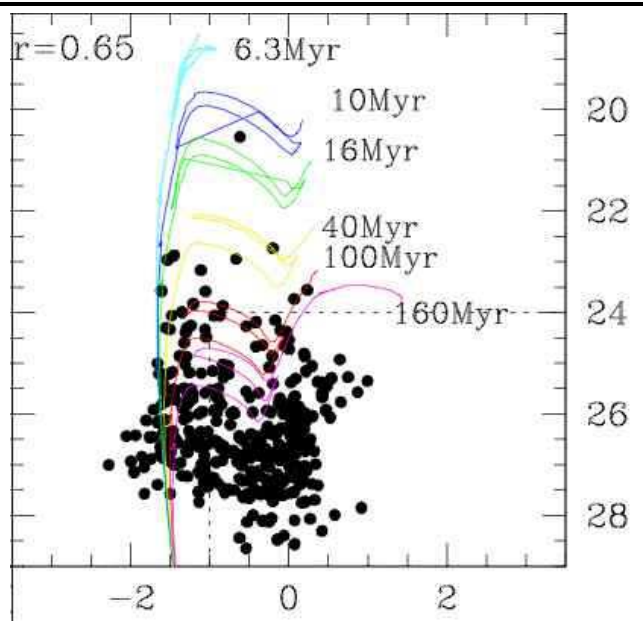}
\caption{CMD of the stellar group shown in Fig. 6} 
\label{cmdnoneb}
\end{figure}

\subsection{Background contaminated cluster}

In Fig. 8 we show a stellar group for which the stellar background
from the host galaxy becomes important. As one can see by comparing
this group (located in the center of the image and identifiable via
the numerous blue stars) with the very compact cluster of unresolved
stars (in the upper left-hand corner), the group is already fairly
spread out within the main body of the galaxy. It is also more
spatially spread out than the two previous young groups. The CMD
(Fig.\,9) clearly shows that the number of stars in the contamination
section of the CMD (see Section 4.1) contributes significantly to the
total number of stars in the group. According to the Padova isochrones
and using the few bright stars found in the group, we estimate the age
of the cluster at about 60\,Myr.

\begin{figure}[t]
\includegraphics[width=\columnwidth]{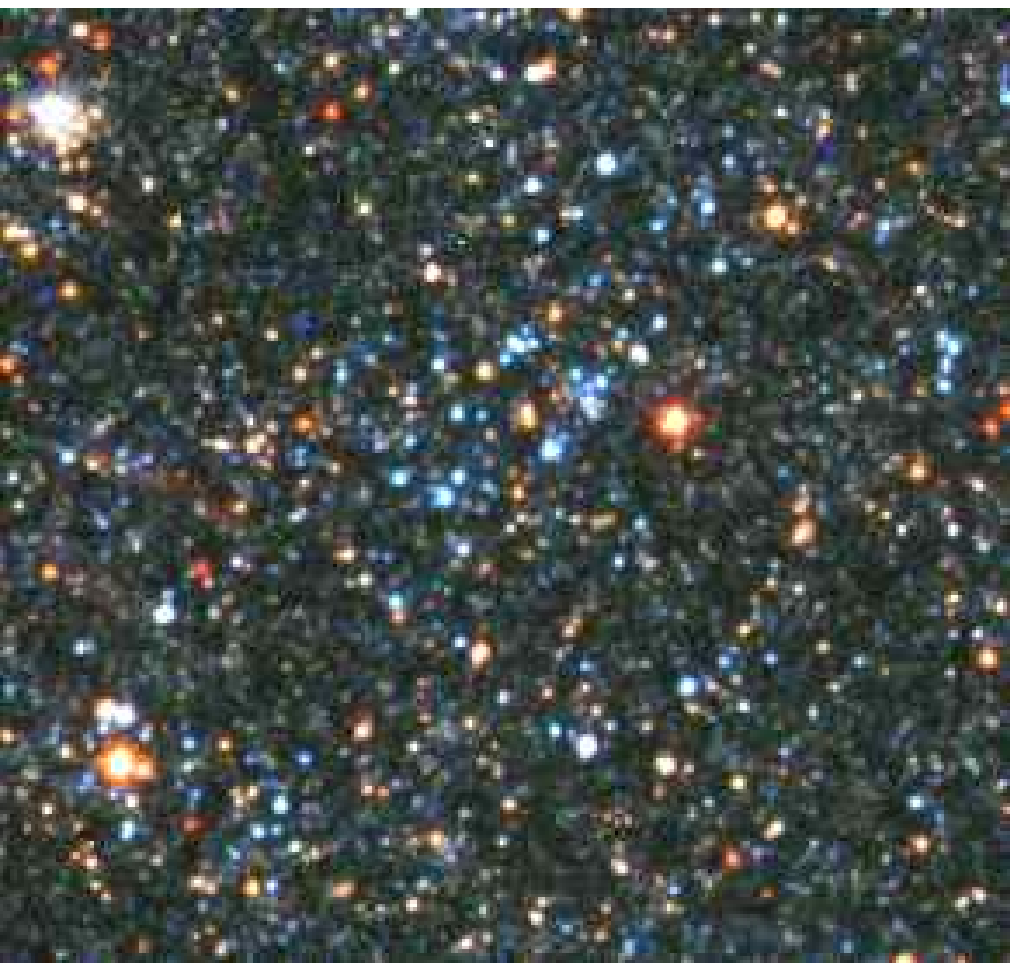}
\caption{Example of a stellar group that is highly contaminated by the
stellar background. The image is about 300\,pc in size}
\label{imconta}
\end{figure}

\begin{figure}[t]
\includegraphics[width=\columnwidth]{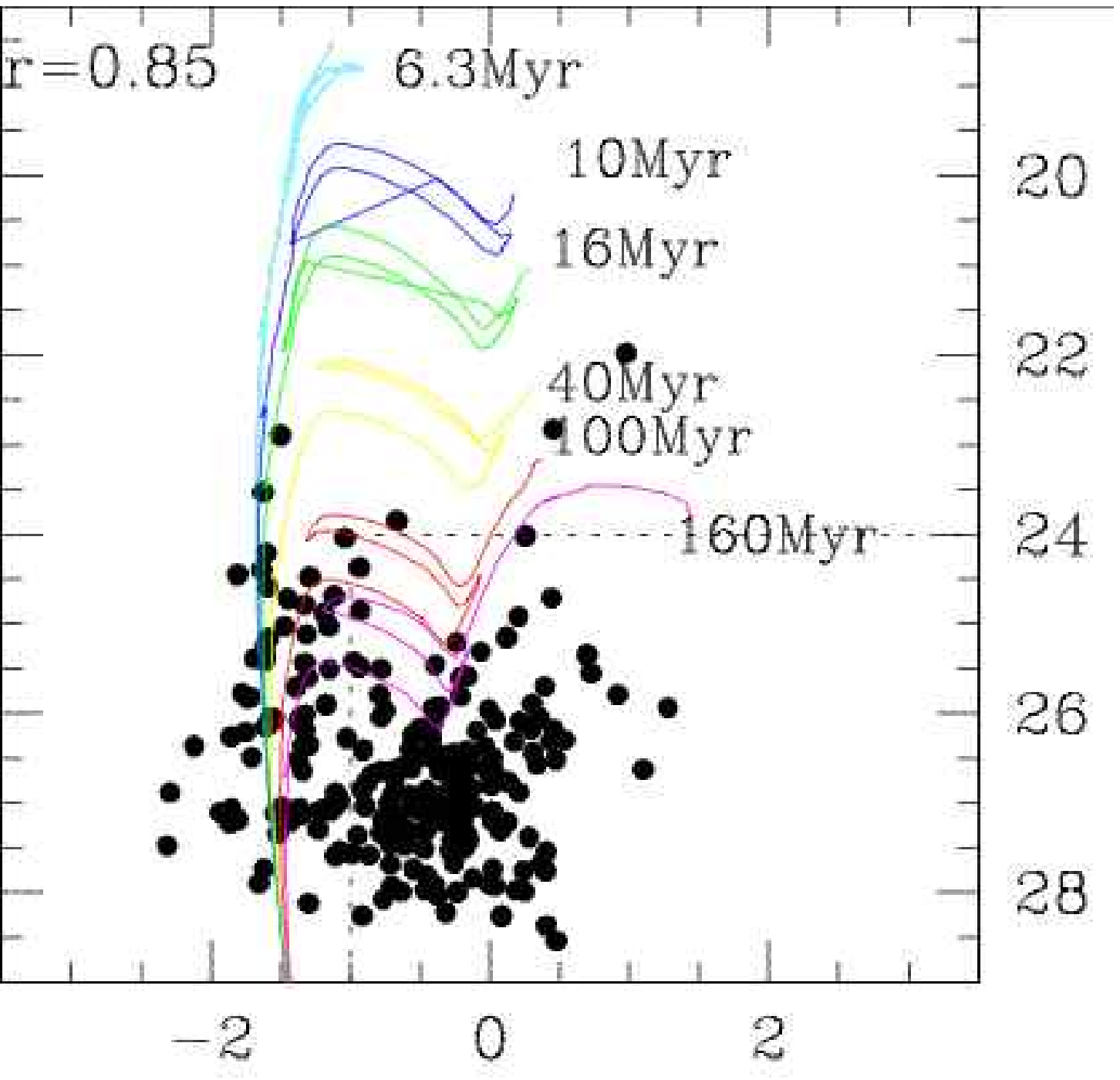}
\caption{CMD of the stellar group shown in Fig.\,8} 
\label{cmdconta}
\end{figure}

\subsection{Dissolved cluster candidate}

The group shown in Fig.\,10 can hardly be called a group. A few
massive stars are observed within a $\sim$$500\times500$\,pc$^2$ field
of view, but it is impossible at this stage to know whether those
stars were born together or if they originate from different parent
clouds. Its CMD (Fig.\,11) shows several stars consistent with an
isochrone of $\sim$100\,Myr. It is, however, interesting to see that
the stars on the 100\,Myr isochrone are well separated from the groups
of red evolved stars seen in the contamination section, which suggests
that the group is a good candidate for a dissolved (or nearly
dissolved) star cluster. Note that it would not be possible here to
find groups much older than $\sim$100\,Myr, because this corresponds
to the magnitude cutoff used to select the massive-star sample for the
HOP algorithm.

\begin{figure}[t]
\includegraphics[width=\columnwidth]{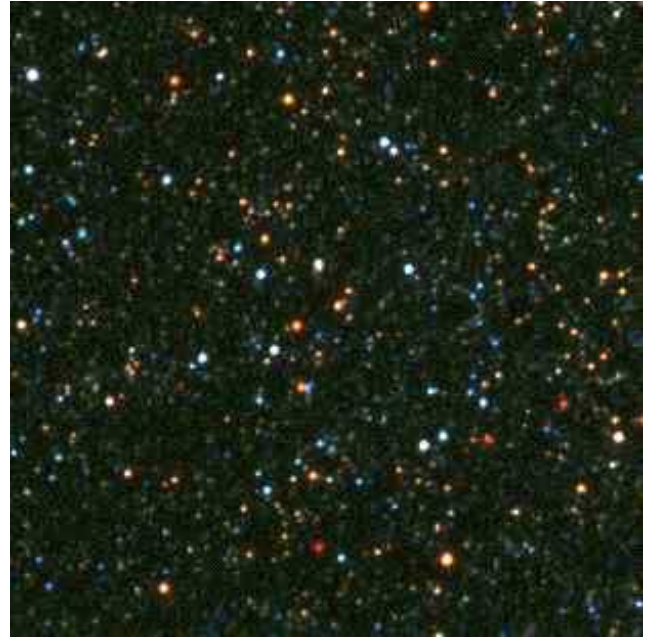}
\caption{Example of a stellar group that is highly contaminated by the stellar
background. The group covers a field of $\sim$500\,pc}
\label{imbckg}
\end{figure}

\begin{figure}[t]
\includegraphics[width=\columnwidth]{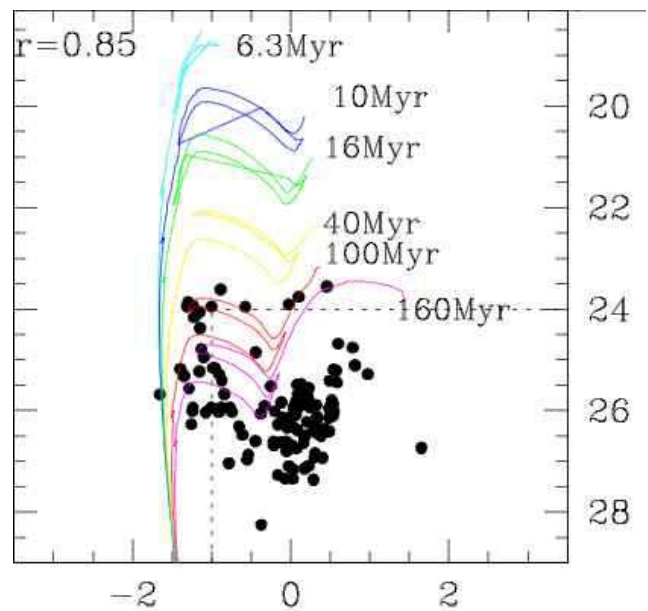}
\caption{CMD of the stellar group shown in Fig.\,10} 
\label{cmdbckg}
\end{figure}

\section{Conclusion and future work}

In this contribution we have presented the very first results of a new
technique to detect and study groups of stars that are very dispersed
and contaminated by the stellar background. The technique uses the
photometric data of individual stars and a clustering algorithm to
find groups of stars. CMDs are then used to study the physical
properties of the groups, such as their age.

We tested our technique on {\sl HST}/ACS images of the spiral galaxy
IC2574. We presented four different examples from among the 79 groups
found so far, and estimated their ages to range from $\leq$10\,Myr to
$\sim$100\,Myr. Of these, 11 groups found in this study are good
candidates for dissolving star clusters.

More physical characteristics will be quantified in the future for the
clusters found in IC2574, including the size of the groups, their
density, shape, masses and compactness. Eventually, more clusters
will be found in many other galaxies. Their characteristics will be
studied as a function of various properties of the host galaxy, such
as the radial location within the host galaxy, the star-formation
intensity, the morphological type, and the potential well. Hopefully,
this will lead to a better understanding of the evolution of star
clusters and the role of the host galaxy in their destruction and
dissolution.

%
 \acknowledgments
This work was supported by {\sl HST} grant HST-AR-10968.02A.


%

\begin{thebibliography}{}
 
\bibitem[\protect\citeauthoryear{Bastian \& Goodwin}{2006}]{bas06} 
Bastian, N. \& Goodwin, S. P., 2006, MNRAS, 369L, 9

\bibitem[\protect\citeauthoryear{Eisenstein \& Hut}{1998}]{hop} Eisenstein, D. J. \& Hut, P. 1998, ApJ, 498, 137

\bibitem[\protect\citeauthoryear{Fall et al.}{2005}]{fall05} Fall, S. M., Chandar, R., \& Whitmore, B. C. 2005, ApJ, 631, L136

\bibitem[\protect\citeauthoryear{Girardi}{2006}]{gir06} Girardi 2006, private communication [http://pleiadi.pd.astro.it/]

\bibitem[\protect\citeauthoryear{Karachentseva et al.}{2003}]{kara03} Karachentsev, I. D., Makarov, D. I., Sharina, M. E., Dolphin, A. E., Grebel, E. K., Geisler, D., Guhathakurta, P., Hodge, P. W., Karachentseva, V. E., Sarajedini, A., \& Seitzer, P. 2003, A\&A, 398, 479

\bibitem[\protect\citeauthoryear{Lada \& Lada}{2003}]{lada03} Lada, C. J. \& Lada, E. A. 2003, ARA\&A, 41, 57

\bibitem[\protect\citeauthoryear{Pellerin et al.}{2007}]{pel07} Pellerin, A., Meyer, M. J., Harris, J., \& Calzetti, D. 2007, ApJ, 658, L87

\bibitem[]{} Schlegel, D. J., Finkbeiner, D. P., \& Davis, M. 1998, ApJ, 500, 525

\bibitem[\protect\citeauthoryear{Sirianni et al.}{2005}]{sir05} Sirianni, M., et al. 2005, PASP, 117, 1049

\end{thebibliography}

%

\end{document}